\documentclass[apj]{emulateapj}
\usepackage{apjfonts}
\usepackage{natbib}
\usepackage{graphicx}
\bibpunct{(}{)}{;}{a}{}{,}
\newcommand{\kms}{{\rm km~s$^{-1}$}}

\begin{document}
\title{
High-resolution proper motions 
in a sunspot penumbra
}
   \author{I.~M\'arquez\altaffilmark{1,2},
   	J.~S\'anchez~Almeida\altaffilmark{1},
	J.~A.~Bonet\altaffilmark{1}}
    \altaffiltext{1}{Instituto de Astrof\'\i sica de Canarias, 
              E-38205 La Laguna, Tenerife, Spain}
    \altaffiltext{2}{Departamento de An\'alisis
     	Matem\'atico, Universidad de La Laguna, E-38271 La Laguna, 
     	Tenerife, Spain}
   \email{imr@iac.es,jos@iac.es,jab@iac.es}
\begin{abstract}
Local correlation tracking techniques
are used to
measure proper motions in a series of high 
angular resolution ($\sim$0\farcs 1) penumbra images.
If these motions trace true plasma motions, then
we have detected converging flows that 
arrange the plasma in long narrow filaments
co-spatial with dark penumbral filaments.
Assuming that these flows are 
stationary, the vertical 
stratification of the atmosphere and
the conservation of mass 
suggest
downflows 
in the filaments of the order of 200~m~s$^{-1}$.
The
association between downflows and dark
features may be a sign of convection,
as it happens with the non-magnetic granulation.
Insufficient spatial resolution may explain why
the estimated vertical velocities are not
fast enough to supply the radiative losses 
of penumbrae.
\end{abstract}

%



\section{Introduction}\label{intro}
Sunspot penumbrae appear
in the first telescopic observations of
sunspots made four hundred years ago
\citep[see the historical introduction by][]{cas97}.
Despite this long observational record, we
still lack of a physically
consistent scenario to explain their structure, origin and 
nature. Penumbrae are  probably  a form of convection
taking place in
highly inclined strong magnetic fields
\citep{dan61,tho93,sch98a,hur00,wei04}. 
However, there is no  consensus even on this
general description. For example, the {\em observed}
vertical velocities do not suffice to
transport the energy radiated away by penumbrae
\citep[e.g.,][and \S~\ref{vertical}]{spr87},
which has been used to argue that they are not exclusively
a convective phenomenon. The
difficulties of understanding and modeling
penumbrae are almost certainly associated with the
small length scale at which the relevant
physical process take place. This limitation
biases all observational descriptions, and it also 
makes the numerical modeling challenging and
uncertain.

From an observational point of view, one approaches the
problem of resolving the physically interesting
scales by two means. First, 
assuming the existence of unresolved
structure when analyzing the data,
 in particular,
when interpreting
the spectral line asymmetries 
\citep[e.g.,][]{bum60,gri72,gol74,san92b,wie95,
sol93b,bel04,san04b}.
Via line-fitting, and with a proper modeling, this 
indirect technique allows us to infer physical
properties of unresolved structures.
On the other hand, one gains spatial
resolution by directly  improving the image quality of the 
observations, which involves both 
the optical quality of the instrumentation 
and the application of  image 
restoration techniques
\citep[e.g., ][]
	{mul73a,mul73b,bon82,bon04b,bon04,sta83,lit90b,
	tit93,sut01b,sch02,rim04}.
Eventually, the two approaches
have to be combined  when the relevant length-scales
are comparable to the photon mean-free-path 
\citep[see, e.g.,][]{san01b}.
The advent of the  Swedish Solar Telescope
\citep[SST; ][]{sch03c,sch03d} has opened up  new possibilities
along the second direct course. Equipped with adaptive optics (AO), it  
allows us
to revisit old unsettled issues with unprecedented 
spatial resolution ($\sim$0\farcs 1),
a strategy
which often brings up 
new observational results. In this sense the
SST  has already  
discovered a new penumbral structure,
namely, dark lanes flanked 
by two bright filaments \citep{sch02,rou04}. These
dark cores in penumbral filaments 
were neither expected nor theoretically predicted, which
reflects the gap between our understanding
and the penumbral phenomenon.

The purpose of this work is to describe
yet another new finding arising from 
SST observations of penumbrae. It turns out that
 the penumbral proper motions
diverge away from bright filaments and
 converge toward dark penumbral filaments.
We compute the proper motion
velocity field employing  the local
correlation tracking method 
(LCT) described in the next section. 
Using the mean velocity field computed in this
way, we follow the evolution of a set of 
tracers (corks) 
passively advected by the mean velocities.
Independently of the details of this computation,
the corks tend to form long narrow filaments 
that avoid the presence of bright filaments (Fig.~\ref{cork1}b).
This
is the central finding of the paper, whose details 
uncertainties and consequences are discussed in the 
forthcoming sections.
The behavior resembles  the
flows in the non-magnetic Sun associated with the 
granulation, mesogranulation,
and supergranulation
\citep[e.g.,][]{nov89,tit89,hir97,wan95,ber98}.
The matter at the base of the photosphere
moves horizontally from the sources of uprising plasma to the 
sinks in cold downflows. 
Using this 
resemblance to granular convection,
we argue that the observed proper motions
seem to indicate
the existence of downward motions
throughout penumbrae, and in doing so, they 
suggest  the convective nature
of the penumbral phenomenon.

LCT 
techniques have been applied to penumbrae
observed with lower spatial resolution 
 \citep[see][]{wan92,den98}.
Our analysis confirms previous findings of
radial motions 
inward or outward depending on the distance
to the penumbral border. In addition, we discover 
the convergence of these radial flows to form
long coherent filaments.   

The paper is organized as follows.
The  observations and data analysis are summarized in
\S~\ref{observations}. The proper motions
of the small scale penumbral features are
discussed in \S~\ref{results} and \S~\ref{formation}. 
The vertical
velocities to be expected if the proper 
motions trace true mass motions are
discussed
in \S~\ref{vertical},
where we also consider their potential for
convective  transport in penumbrae.
Finally, we elaborate on the implications
of our finding in \S~\ref{conclusions}.  
The (in-)dependence of the results on 
details of the algorithm is analyzed in
Appendix~\ref{robust}.

\section{Observations and data analysis}\label{observations}

We employ 
the original data set of \citet{sch02}, generously
offered for public use by the authors.
They were obtained with the SST \citep{sch03c,sch03d},
a refractor with 
a primary lens of  0.97~m and  equipped with
AO. The data
were post-processed to render
images near the diffraction limit.
Specifically, we  study the behavior of a penumbra
in a 28 minutes long sequence
with a cadence of 22 s between snapshots. The penumbra
belongs to the large
sunspot of the  active region NOAA 10030, 
observed on July 15, 2002, close to the
solar disk center (16\degr\ heliocentric angle).
The series was 
processed with Joint Phase-Diverse Speckle 
\citep[see][]{lof02},
which provides
an angular resolution of 0\farcs12,
close to the
diffraction limit of the telescope 
at the working wavelength (G-band,  $\lambda$~4305~\AA).
The field-of-view (FOV) is $26\arcsec\,\times\,40\arcsec$, 
with pixels 0\farcs041 square.
The images of the series were corrected for diurnal
field rotation, rigid aligned,
destretched, and subsonic Fourier filtered\footnote{The
 subsonic filter
removes fast oscillations
mostly due to p-modes
 and residual 
jitters stemming from  destretching
\citep{tit89}.
}
(modulations larger than 4 km~s$^{-1}$  are suppressed).
For additional details, see \citet{sch02} and 
the web page set up to distribute  the  data\footnote{
\url{http://www.solarphysics.kva.se/data/lp02/}}. 
\begin{figure*}
\includegraphics[angle=90,width=18cm]{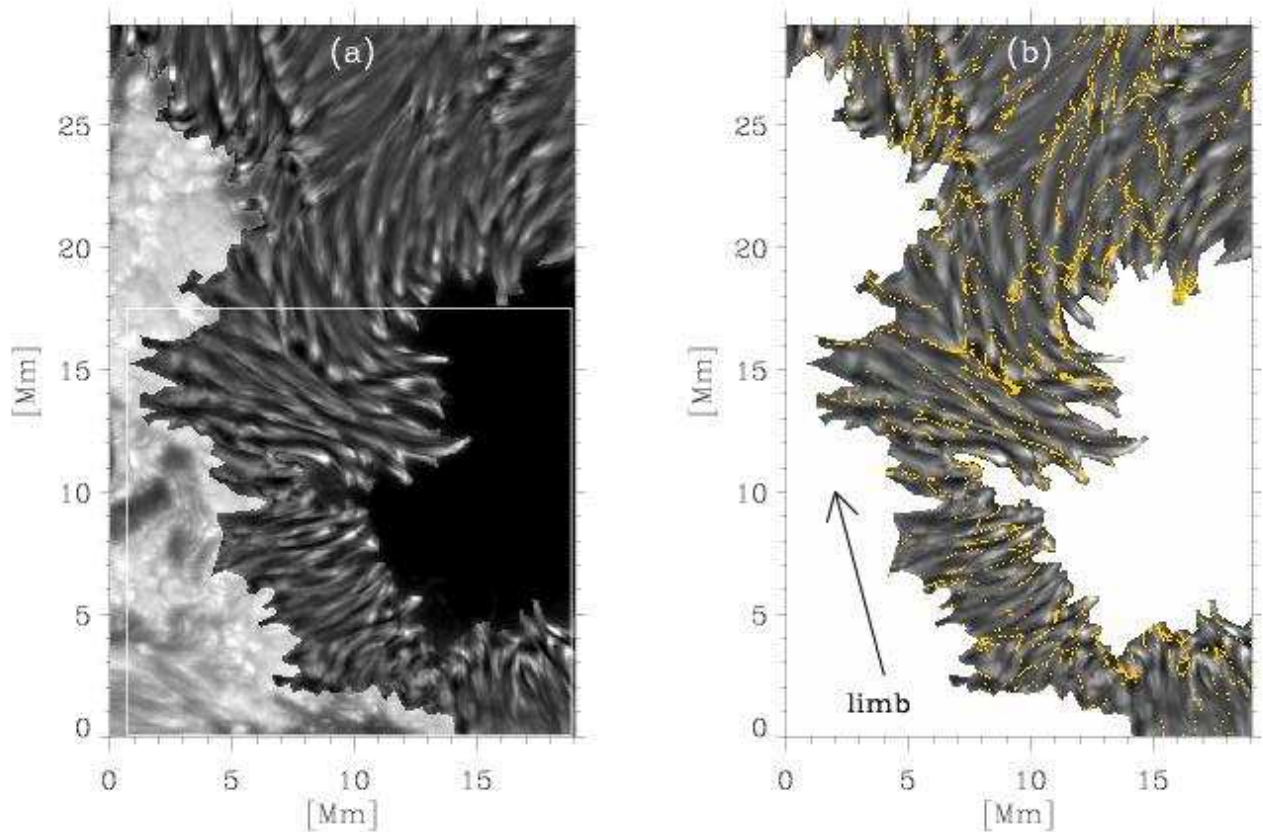}
\caption{(a) Mean image of the penumbra 
sharpened by removal of a 
spatial
running mean of the original
image. The figure also includes
the umbra  
and sunspot surroundings for reference.
(b) Image with the corks  remaining in the penumbra 
after 110~min (the yellow dots).
The corks form long narrow filaments that
avoid the bright penumbral 
filaments (compare the two figures).
The large arrow points out the direction of the closest 
solar limb.
}
\label{cork1}
\end{figure*}

Our work is devoted  to the penumbra,
a region which we select by visual inspection
of the images. Figure~\ref{cork1}a shows the
full FOV with the penumbra artificially
enhanced with respect of the  
umbra and the surrounding sunspot moat. Figure~\ref{cork1}b
shows the penumbra alone. The qualitative
analysis carried out in the next sections
refers to the lower half of the penumbra,
enclosed by a white box in Fig.~\ref{cork1}a.
The results that we describe are more
pronounced in here, perhaps, because
the region is not under the influence of a
neighbor penumbra outside but close to
the upper part of our FOV. 
The effects of considering the full FOV
are studied in Appendix~\ref{robust}.

We compute  proper motions  
using the local correlation tracking 
algorithm (LCT) of \citet{nov88}, as implemented
by \citet{mol94}.
It works by selecting small sub-images around the same
pixel in contiguous snapshots that are cross-correlated
to find the displacement of best match.
The procedure provides a  
displacement or proper motion per
time step, which we average in time. These mean displacements
give the (mean) proper motions  analyzed in the work.
The sub-images are defined using a 
2D 
Gaussian window with
smooth edges.
The size of the window must be set according to the 
size of the structures selected as tracers.
As a rule of thumb,
the size of the window is half the size of the structure
that is being tracked (see, e.g.,  \citealt{bon04}).
We adopt a window of FWHM  5~pixels  
($\equiv$~0\farcs 2), 
tracking small features of about 10 pixels
($\equiv$~0\farcs 4).
The LCT algorithm  restricts the relative displacement between 
successive images to a maximum of 2 pixels. This limit constrains 
the reliability of the proper motion velocity 
components, $U_x$ and $U_y$, 
to a maximum of $\pm 2$ pix per time step, 
which corresponds to a maximum velocity of some 3.8~km~s$^{-1}$.
(Here and throughout the paper,
the symbols $U_x$ and $U_y$ 
represent the two Cartesian components of the 
proper motions.)   
Such upper 
limit fits in well the threshold imposed on the original time series,
were motions larger than 4~{\kms\,} were filtered out 
by the subsonic filter.

Using the mean velocity field,
 we track the evolution of passively 
advected tracers (corks) spread out all over the penumbra
at time equals zero. Thus we construct a {\it cork movie}
(not to be confused with the 28~min long sequence of images 
from which the mean velocity field we inferred, which we call
{\em time series}).
The motions are integrated in time assuming 22~s
per time step, i.e., the cadence of the time series. 
Figure~\ref{cork1}b shows the corks that remain in the penumbra
after 110~min. 
(The cork movie can be found in 
\url{http://www.iac.es/proyect/solarhr/pencork.html}.)
Some 30~\% of the original corks leave the
penumbra toward the umbra, the 
photosphere around the sunspot, or the penumbra outside the FOV. 
The remaining 70~\% 
are concentrated in long narrow filaments which
occupy a small fraction of the penumbral area,
since many corks end up in each single pixel.
As it happens with the rest of the free parameters
used to find the filaments, 
the final time  is not critical
and it  was chosen by trial
and error as a compromise that  yields well defined
cork filaments. Shorter times lead to fuzzier filaments,
since the corks do not have enough time to concentrate.  
Longer times erase the filaments because the
corks exit the penumbra or concentrate in a few sparse 
points.

We will compare the position of the cork filaments
with the position of penumbral filaments in the 
intensity images. 
Identifying penumbral filaments is not free from ambiguity,
though.
What is regarded as a filament depends on the spatial resolution
of the observation. (No matter whether
the resolution is 
2\arcsec\  or 0\farcs 1, penumbrae do
show penumbral filaments. Obviously, 
the filaments
appearing with 2\arcsec\  and 0\farcs 1
cannot correspond 
to the same structures.) Moreover,
being dark or bright is a local concept. 
The bright filaments
in a part of the penumbra can be darker than 
the dark filaments elsewhere
\citep[e.g.,][]{gro81}. Keeping in mind these caveats,
we use
the average intensity along the observed time series
to define bright and dark filaments, since it
has the same spatial
resolution as the LCT mean velocity map. In addition,
the local mean intensity of this time average image is removed
by subtraction of a running box mean of the average
image. The removal of low spatial frequencies allows us
to compare bright and dark filaments of
different parts of the penumbra. The
width of the box is set to 41 pixels or 1\farcs 7. 
(This 
time averaged and sharpened intensity  
is the one represented
in Figs.~\ref{cork1}.)

The main trends and correlations to be described
in the next sections
are not very sensitive to
the actual free parameters used to infer them 
(e.g., those defining the LCT and the local  mean intensities).
We have had to choose a particular set 
to provide quantitative estimates,
but the
variations resulting from using other
sets are examined in Appendix~\ref{robust}.

A final clarification may be needed. The 
physical properties of the corks forming
the filaments will be 
characterized using histograms
of physical quantities. We compare
histograms at the beginning of the 
cork movie  with histograms at the
end. Except at time equals zero,
several corks may coincide in a single
pixel. In this case the corks are considered
independently, so that each pixel contributes 
to a histogram as many times as the number
of corks that it contains.

\section{Proper motions}\label{results}

\begin{figure}
\plotone{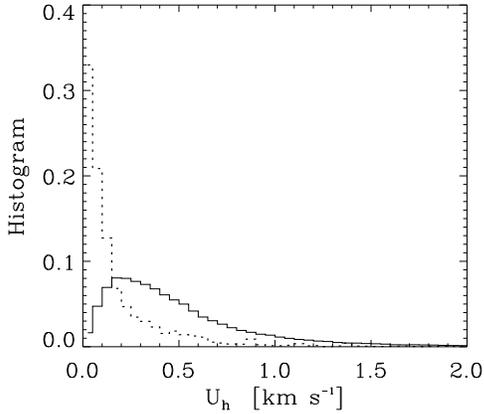}
\caption{Histogram of the horizontal velocities 
in the penumbra as inferred from LCT techniques. The 
solid line shows the whole set of penumbral points.
The 
dotted
line corresponds to the velocities at the
positions of the corks after 110~min.
The corks drift toward low 
horizontal velocity regions.
}
\label{horiz_vel}
\end{figure}

We employ the symbol ${\bf U}_h$
to denote the  proper motion vector.
Its magnitude $U_h$
is given by
\begin{equation}
U_h^2=U_x^2+U_y^2,\label{myuh}
\end{equation} 
with $U_x$ and $U_y$ the two Cartesian 
components.
Note that these components are in a plane
perpendicular to the line of sight. Since the sunspot
is not at the disk center, this plane is not exactly
parallel to the solar surface. However, the differences
are insignificant for the kind of qualitative argumentation
in the paper (see item~\#~\ref{case_coor} in 
Appendix~\ref{robust}). It is assumed that the
plane of the proper motions defines
the solar surface so that the $z$ direction
corresponds to the solar radial direction.
The solid line in 
Figure~\ref{horiz_vel}
shows the histogram
of time-averaged
horizontal velocities considering
all the pixels in the penumbra selected
for analysis (inside the box in Fig.~\ref{cork1}a).
The dotted line of the same figure
corresponds to the distribution of velocities for the
corks that form the filaments (Fig.~\ref{cork1}b).
The typical proper motions in the penumbra are 
of the order of half a \kms . 
Specifically, 
the mean and the standard deviation
of the solid line in Fig.~\ref{horiz_vel}
are 0.51~\kms\ and 0.42~\kms , respectively. With time, the 
corks originally spread all over the
penumbra tend to move toward low horizontal velocities.
The dotted
line in Fig.~\ref{horiz_vel} corresponds to the
cork filaments in Fig.~\ref{cork1}b, and it is characterized
by a mean value of 0.21~\kms and a standard deviation
of 0.35~\kms .
This migration 
of the histogram
toward low velocities is to be
expected since the large proper motions expel
the corks making it 
difficult to form filaments.

\begin{figure}
\plotone{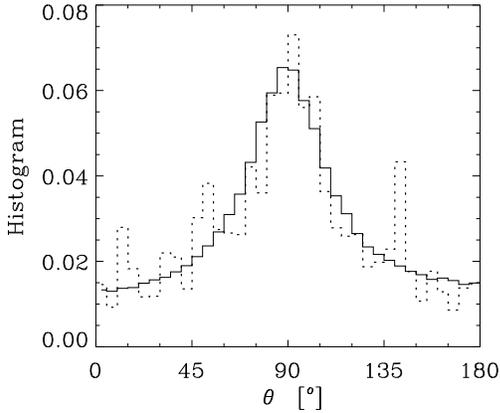}
\caption{Histogram of the angle between the horizontal velocities and
the horizontal gradients of intensity. They tend to be
perpendicular (the histograms peak at  
$\theta\sim$ 90$^\circ$). The same
tendency characterizes both all penumbral points
(the solid
line), and  those having cork filaments (the
dotted  line).
}
\label{angle}
\end{figure}

\begin{figure}
\plotone{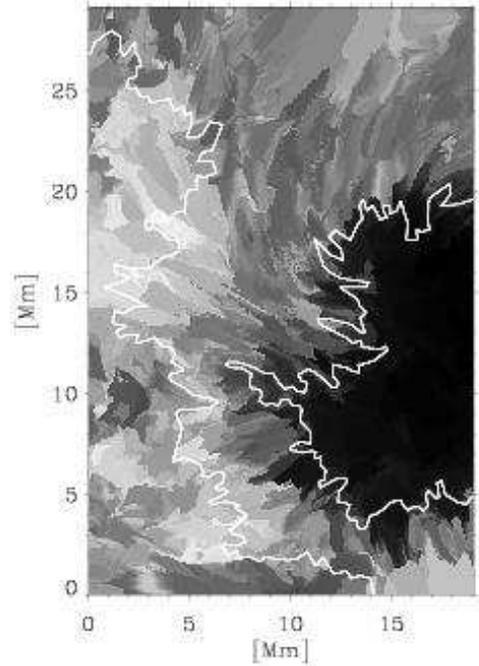}
\caption{Mapping of the final intensities that
corks in each location of the FOV have 
by the end of the cork movie.
The points in the outer penumbra tend to
exit the sunspot,
and so they are bright in this image. The points
in the  inner penumbra
remain inside the sunspot, either within the 
penumbra or the umbra.
The white contours outline  
the boundaries of the penumbra
in Fig.~\ref{cork1}. 
Points outside
the penumbra are also included. 
}
\label{forecast}
\end{figure}

The proper motions are predominantly radial,
i.e., parallel to the bright and dark  penumbral 
filaments traced by the  intensity. 
Figure~\ref{angle} shows the distribution of angles
between the horizontal gradient of 
intensity and the velocity. Using the symbol
$I$ to represent the intensity image, the
horizontal gradient of intensity $\nabla_hI$ is given in 
Cartesian coordinates by, 
\begin{equation}
\nabla_hI=
\Big({{\partial I}\over{\partial x}}~~~~
{{\partial I}\over{\partial y}}\Big)^\dag,
\end{equation} 
with the superscript $\dag$ representing transpose matrix.
The intensity gradients point out the direction perpendicular
to the intensity filaments.
The angle $\theta$  between the
local velocity and the local gradient of intensity 
is 
\begin{equation}
\cos\theta={{\nabla_hI}\over{|\nabla_hI|}}\cdot{{\bf U}_h\over{U_h}}.
	\label{mytheta}
\end{equation}
The angles computed according to the
previous expression tend to be
around 90$^\circ$ (Fig.~\ref{angle}), meaning that the
velocities are perpendicular to the intensity
gradients and so, parallel to the filaments.
The radial motions tend to be inward in the inner penumbra and
outward in the outer penumbra, a systematic
behavior that can be inferred from Fig.~\ref{forecast}. 
At the starting position  
of each cork (i.e., the position at time equals
zero), Fig.~\ref{forecast} shows the intensity  
that the cork reaches by the end cork movie
(i.e., at time equals 110~min).
This forecast 
intensity image has a clear divide in the penumbra.
Those points close enough to the photosphere around the
sunspot
will become bright, meaning that they exit the penumbra.
The rest are  dark implying that they either remain 
in the penumbra or move to the umbra.
Such radial proper motions 
are well known in the penumbral
literature 
\citep[e.g.,][]{mul73a,den98,sob99b}. However, 
on top of this predominantly radial flow, 
there is a
small transverse velocity responsible for the
accumulation of corks in filaments (\S~\ref{formation}).

\begin{figure}
\plotone{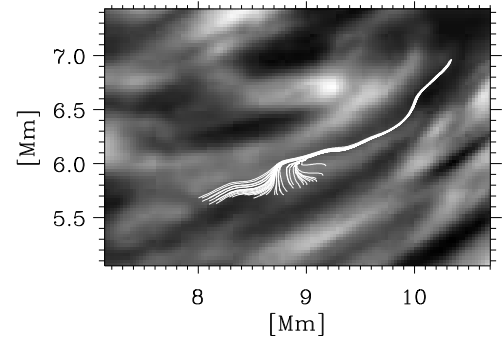}
\caption{Tracks followed by a set of 80
corks that end up in one of the cork filaments.
The spatial coordinates have the same origin
as those in Figs.~\ref{cork1} and \ref{forecast}.
}
\label{tracks}
\end{figure}

\begin{figure}
\plotone{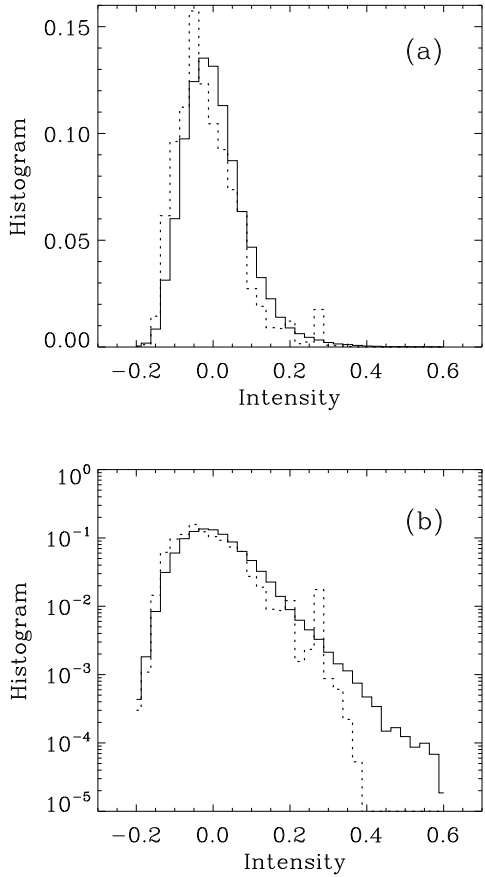}
\caption{(a) Histograms of the distribution of intensity
associated with the cork filaments. The original distribution
corresponds to corks uniformly spread out 
throughout the penumbra (the solid line).
The 
dotted
line represents the distribution at the
cork filaments. Note the  shift.  All  intensities are referred
to the local mean intensity, which explains the
existence of negative
abscissae.
(b) Same as (a) but in logarithm scale to appreciate
the lack of very bright features associated with the 
cork filaments.
}
\label{histo1}
\end{figure}

The corks in Fig.~\ref{cork1}b
form long chains that avoid bright
filaments and overlie dark filaments. 
The tracks followed by a set of corks finishing up in one 
of the filaments are plotted in Fig.~\ref{tracks}.
It shows both the gathering at the head of the filament, and
the tendency to avoid bright features 
where the narrow cork filament is formed.
The migration of the corks toward dark penumbral filaments
is quantified  in 
Fig.~\ref{histo1}a. It
shows the histogram of intensities associated
with the initially uniform distribution of corks throughout
the penumbra (the solid line), and 
the final histogram after 110~min (the 
dotted line). A global
shift toward dark locations is clear. The change is of the
order of 20\%, as defined by,
\begin{equation}
\Delta\mu/\sigma\equiv\big[\mu_I(110)-\mu_I(0)\big]/\sigma_I(0)\simeq -0.21,
\label{shift}
\end{equation}
with $\mu_I(t)$ the mean and
$\sigma_I(t)$ the standard deviation
of the histogram of intensities at time $t$~min.
Figure~\ref{histo1}b contains the same histogram
as Fig.~\ref{histo1}a but in
a logarithm scale. It allows us to appreciate 
how the shift of the histogram
is particularly enhanced in the tail of large intensities.
The 
displacement between the two histograms
is not larger 
because the corks do not end up in the darkest parts
of the dark penumbral filaments (see, e.g., Fig.~\ref{tracks}).

\section{Formation of  cork filaments}\label{formation}

It is important to know which 
properties of the velocity field 
produce  the formation of filaments. 
Most cork filaments are only a few pixels wide (say, from 1 to 3).
The filaments are so narrow that they seem to trace
particular  stream lines,
i.e., the 1D path followed by a test particle fed at the 
starting point of the filaments. Then the
presence of a filament requires both 
a low value
for the velocity in the filament, and 
a continuous source
of corks at the starting point. 
The first property 
avoids the evacuation of the filament
during the time span of the cork
movie, and it is assessed
by the results in \S~\ref{results},
Fig.~\ref{horiz_vel}. The second
point  allows the flow
to collect the many corks that trace each filament
(a single cork cannot outline a filament.)
If the cork filaments are formed in this way, then 
their widths are independent of the LCT window 
width.

\begin{figure}
\plotone{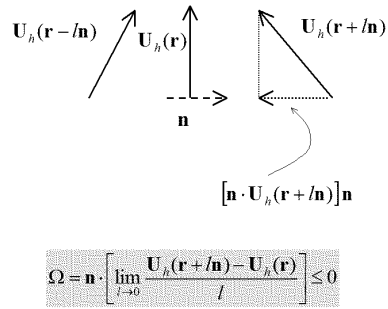}
\caption{The solid line vectors show three velocities
of  a convergent 
velocity field. The 
dashed line vector {\bf n} corresponds to the 
vector normal to the velocity
${\bf U}_h$ in the point {\bf r}, so that  {\bf n}$\cdot {\bf U}_h({\bf r})=0$.
The dotted line vector shows the component of the velocity field
parallel to {\bf n} when moving along {\bf n}. Note
that it is anti-parallel to {\bf n}, which implies that the
directional gradient $\Omega$ is negative for convergent
field lines.}
\label{topo1}
\end{figure}

\begin{figure}
\plotone{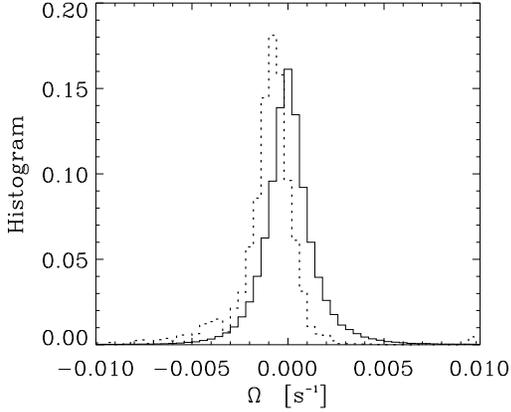}
\caption{Histograms of the derivative of the
velocities in the direction perpendicular 
to the velocity field ($\Omega$; see its 
definition in the main text).
As in  previous figures, the solid line
corresponds to all the pixels in the
penumbra whereas the dotted
line describes
the cork filaments.}
\label{topo2}
\end{figure}

For the corks to gather, the  stream
lines  of different corks have to converge.
A convergent velocity field has the topology
shown by the three solid line vectors in Fig.~\ref{topo1}.
As it is represented in the figure, the spatial
variation of the velocity 
vector ${\bf U}_h$ in the direction {\bf n} perpendicular
to ${\bf U}_h$  is
anti-parallel to ${\bf n}$. (${\bf U}_h\cdot {\bf n}=0$ with
 $|{\bf n}|=1$.) Consequently, the places
where the velocities converge are those where 
$\Omega < 0$, with 
\begin{equation}
\Omega={\bf n}\cdot[({\bf n}\nabla_h){\bf U}_h]. 
\label{omega}
\end{equation}
The equation  follows from the expression for
the variation of a vector ${\bf U}_h$
in the direction of the vector {\bf n}, which is given by
$({\bf n}\nabla_h){\bf U}_h$ \citep[e.g., ][ \S~4.2.2.8]{bro85}.
Then the component of this directional derivative in the direction
normal to ${\bf U}_h$ is given by $\Omega$ in equation~(\ref{omega}).
The more negative $\Omega$ 
the larger the convergence rate of the velocity
field.
(Note that the arbitrary sign
of {\bf n} does not affect $\Omega$.)
Using Cartesian coordinates in a plane, equation~(\ref{omega})
turns out to be, 
\begin{equation}
\Omega={{U_x^2}\over{U_h^2}}{{\partial U_y}\over{\partial y}}+
{{U_y^2}\over{U_h^2}}{{\partial U_x}\over{\partial x}}-
{{U_x U_y}\over{U_h^2}}\Big[{{\partial U_y}\over{\partial x}}
+{{\partial U_x}\over{\partial y}}\Big].
\end{equation}
The histograms of $\Omega$ for 
all the pixels in the penumbra and 
for  the cork filaments
are shown in Fig.~\ref{topo2}. 
Convergent and divergent flows coexist in the penumbra
to give a mean $\Omega$ close to zero (the solid line
is characterized by a mean  
of $4\times 10^{-5}$~s$^{-1}$ 
and a standard deviation of $1.8\times 10^{-3}$~s$^{-1}$). However,
the cork filaments trace converging flows
(the  dotted
line has a mean of  
$-1.1\times 10^{-3}$~s$^{-1}$
and a standard deviation of $2.7\times 10^{-3}$~s$^{-1}$).
The typical $\Omega$ at the corks
implies moderate convergent
velocities, of the order of 100~m~s$^{-1}$ for
points separated by 100~km. 
\section{
Discussion
}\label{vertical}
Assume that
the observed proper motions trace
true mass motions.
The horizontal velocities should be
accompanied by vertical motions. 
In particular, those places traced by the cork filaments
tend to collect mass that must be transported
out of the observable layers by vertical
motions. The need for mass conservation
allows us to estimate the magnitude of such vertical velocities.
Mass conservation in a stationary fluid
implies that the divergence of the velocity field times the
density is zero. This constraint leads to,
\begin{equation}
U_z\simeq h_z \Big[{{\partial U_x}\over{\partial x}}+
{{\partial U_y}\over{\partial y}}\Big],
	\label{myuz}
\end{equation}
as proposed by \citet{nov87,nov89}.
A detailed derivation of the equation
is given in Appendix~\ref{appb}.
The symbol $h_z$ stands for the 
scale height of the flux of mass, 
which must be close to the density
scale height \citep[see][]{nov87,nov89}.
We want to stress that 
neither
equation~(\ref{myuz})
assumes the plasma to be 
incompressible, nor
$h_z$ is the scale height of $U_z$.
Actually, the value of 
$h_z$, including its sign,
is mostly set by the vertical 
stratification of density
in the atmosphere
(see Appendix B). 
Figure~\ref{usubz} shows histograms of  $U_z$ computed
using equation~(\ref{myuz}) with $h_z=100$~km.
We adopt this scale height  because
it is close to, but smaller than, the figure measured
in the non-magnetic Sun by \citet[][150~km]{nov89}.
The density scale hight decreases with temperature,
which reduces the penumbral value with respect
to that in the non-magnetic photosphere. 
(One can readily change to any other $h_z$
since it scales all vertical velocities.)
According to Fig.~\ref{usubz}, we find no
preferred upflows or downflows in the penumbra.
The solid line represents the histogram
of $U_z$ considering all penumbral points; 
it has a mean of only $3\times 10^{-3}$~\kms\ 
with a standard deviation of 0.39~\kms . 
However,  the cork filaments prefer
downflows. The dotted line
shows the histogram of $U_z$ for the corks
at the cork filaments.
It has a mean of 
-0.20~\kms  with a standard deviation  of 0.36~\kms .

According to the arguments given 
above, the
cork filaments seem to be  associated
with downflows.
The cork filaments are also associated with
dark features (\S~\ref{results}).
This combination characterizes 
the non-magnetic granulation \citep[e.g.][]{spr90}, and it
reflects the presence of
convective motions.
The question arises as to whether
the velocities that we infer can transport the
radiative flux escaping from penumbrae.
Back-of-the-envelope
estimates yield the following relationship 
between convective energy flux $F_c$, 
mass density $\rho$, specific
heat at constant pressure $c_P$, and
temperature difference 
between upflows  and downflows $\Delta T$,
\begin{equation}
F_c\simeq \rho|U_z|c_P\Delta T
\label{ctransport}
\end{equation}
\citep{spr87}.
Following the arguments by 
\citet[][ \S~3.5]{spr87},
the penumbral densities and temperature differences
are  similar to those
observed in the quiet Sun. Furthermore  $F_c$ is a large
fraction of the quiet Sun flux 
(75~\%). Then the vertical velocities required to account for the
penumbral radiative losses are  of the order of the
velocities in the non-magnetic granulation or 1~km~s$^{-1}$
\citep[see also][]{sch03f,san04c}. 
One may argue that the penumbral vertical
velocities inferred above are far too small to comply
with such needs.
However, the spatial resolution of our velocity
maps is limited. 
The LCT detects the proper motions of structures 
twice the size of the window, or 0\farcs 4 
in our case
(see \S~\ref{observations}).
The  LCT
smears the horizontal velocities and by doing so, it 
smears the vertical velocities too\footnote{Equation~(\ref{myuz})
is linear in $U_x$ and $U_y$ so that it also holds
for averages of $U_x$ and $U_y$, rendering
averages of $U_z$; see Appendix~\ref{appb}}.
Could this bias mask large vertical convective motions?
We believe that it can, as inferred from the
following argument.
When the LCT procedure is applied to normal granulation,
it leads to vertical velocities of a few hundred
m~s$^{-1}$, which are smaller than the fiducial
figure required
for equation~(\ref{ctransport}) to account for the
quiet Sun radiative flux (1~km~s$^{-1}$). However,
the quiet Sun radiative flux is deposited in the 
photosphere by convective motions. 
Therefore a large bias affects the quiet Sun estimates
of vertical velocities based on LCT, and it is 
reasonable 
to conjecture that the same bias also affects 
our penumbral estimates.
The vertical velocities associated with the
non-magnetic solar granulation 
mentioned above have not been obtained
from the literature. We have not been able
to find any estimate when the LCT window 
has a size to track individual granules,
say, 0\farcs 7--0\farcs 8\footnote{\citet{nov89} and \citet{hir97} 
employ a  
larger window to select mesogranulation, whereas 
\citet{wan95b} do not provide units for the
divergence of the horizontal velocities.}. Then, we carried out
an add hoc estimate
using  the quiet region outside the sunspot
studied by \citet{bon04}. The vertical velocities
are computed  employing 
equation~(\ref{myuz}) when the FWHM of the LCT window
equals 0\farcs75 . The inferred vertical
velocities \footnote{The finding of low vertical
velocities should not depend
on the specific observation
we use. The order of magnitude 
estimate, equal for all observations,
leads to 100~m~s$^{-1}$ -- consider a
gradient of horizontal velocities of the order of
1~\kms\ across a granule 1000~km wide. 
Equation~[\ref{myuz}] with $h_z=100$~km renders 0.1~km~s$^{-1}$.} 
have standard deviations between 
280~m~s$^{-1}$ and 110~m~s$^{-1}$ depending of the
time average used to compute the 
mean velocities (5~min and 120~min, respectively).
One can conclude that the vertical velocities
are some five times
smaller that the fiducial 1~km~s$^{-1}$.
This bias would
increase our velocities to 
values consistent with the radiative
losses  of penumbrae.

\begin{figure} 
\plotone{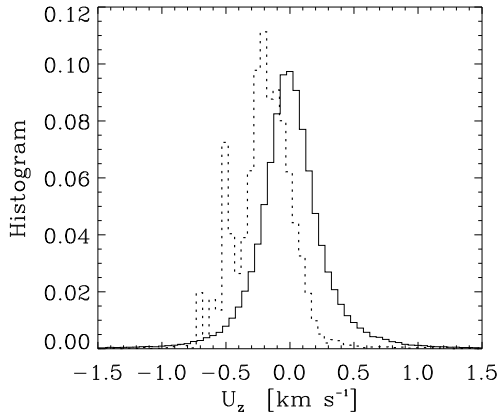}
\caption{Histogram of the vertical velocities 
of the all the pixels in the penumbra (the solid line) and
only those in the cork
filaments (the dotted line).
Upflows yield positive $U_z$.
The corks seem
to be preferentially associated with downflows of a few
hundred m~s$^{-1}$. 
}
\label{usubz}
\end{figure}
In short, given the limited spatial resolution, 
the velocities inferred from LCT
may be underestimating the true 
vertical velocities.
If this bias is similar to that affecting
the non-magnetic granulation, then 
the observed velocities
suffice to transport the radiative losses 
of penumbrae by convection.

All the discussion above assumes the proper motions
to trace true motions.
However, the proper motion velocities 
disagree
with the plasma motions inferred from the
Evershed effect, i.e., the intense and
predominantly  radial outward flows 
deduced from Doppler shifts 
\citep[e.g.,][]{sol03,tho04}. Our proper
motions are both inward and outward, 
and only of moderate speed (\S~\ref{results}).
This disagreement may cast doubts on
the vertical velocities  computed above.
Fortunately, the doubts can be cleared up by   
acknowledging 
the existence of horizontal motions
not revealed by the LCT technique,
and then  working out the consequences.
Equation~(\ref{myuz}) is linear so that different 
velocity components 
contribute separately to $U_z$. In particular,
the Evershed  flow represents
one more component, and it has to be added to the
proper motion based vertical velocities. 
The properties of the Evershed flow 
at small spatial scales and large spatial scales  
do not modify the velocities
in Figure~(\ref{usubz}).
On the one hand, we apply equation~(\ref{myuz}) to average 
proper motions as inferred with the finite spatial resolution of 
our observations. It is a valid approach  which, however, only
provides average vertical velocities (see Appendix~\ref{appb}).
One has to consider the contribution of the 
average Evershed velocities, eliminating 
structures smaller than the spatial resolution of 
the proper motion measurements.
On the other hand, the large scale structure of the 
Evershed flow is also inconsequential. 
According to equation~(\ref{myuz}),
a fast but large spatial scale flow does not 
modify $U_z$;
add a constant to $U_x$ and $U_y$, and $U_z$ 
does not change. 
Only the structure of the Evershed flow at intermediate
spatial scales needs to be considered, and
it does not invalidate our conclusion of
downflows associated with the cork filaments.
For the Evershed flow to invalidate this association, 
it would have to provide upflows co-spatial
with the cork filaments, and so, with
dark lanes. However, the existence of 
upflows in dark lanes is not favored
by the observations of the Evershed effect,
which seem to show the opposite, i.e.,
a local correlation between upflows and 
bright lanes \citep[e.g.,][]{bec69c,san93b,joh93,sch00b}.
Thus we cannot  rule out
a bias of the vertical velocities in Figure~(\ref{usubz})
due to the Evershed flow
but, if existing, the 
observed vertical component of the
Evershed flow seems to
reinforce rather than invalidate the 
relationship
between cork filaments and downflows.

\section{Conclusions}\label{conclusions}
Using local correlation tracking (LCT) techniques, 
we measure mean proper motions in 
a series of high angular resolution
(0\farcs 12) penumbral images
obtained with the 1-meter Swedish Solar Telescope \citep[SST;][]{sch02}.
Previous studies of lower resolution find
predominantly radial proper motions, a result
that we confirm. On top of this 
trend, however, we discover
the convergence of the radial flows to form
long coherent filaments.    
Motions diverge away from bright filaments to
converge toward dark filaments.
The behavior resembles  the
flows in the non-magnetic Sun associated with the 
granulation, where
the matter moves horizontally from the sources
of uprising plasma to the 
sinks in cold downflows. 
Using such similarity,
we argue that the observed proper motions
suggest the existence of downward  flows
throughout the penumbra,  
and so, they suggest the convective nature
of the penumbral phenomenon. 
The places where the proper motions converge would
mark sinks in the penumbral convective pattern.

The presence of this convergent motions is best
evidenced using tracers passively advected
as prescribed by the penumbral proper motion
velocity field: see the dots in Fig.~\ref{cork1}b
and Fig.~\ref{tracks}. With time,
these tracers or corks form
filaments that avoid the bright features 
and tend to coincide with  dark structures. We quantify
this tendency by  following the time evolution
of corks originally spread throughout the
penumbra. After 110~min, the corks overlie
features which are significantly fainter than the
mean penumbra (the histogram of intensities is
shifted by 20\%; see \S~\ref{results}). 
Assuming that the proper motions
reflect true  stationary
plasma motions, the need for mass conservation
allows us 
to estimate the vertical
velocities at the cork filaments, i.e.,
in those places where the plasma 
converges.
These vertical velocities tend to be directed
downward with a mean of the order of 200~m~s$^{-1}$.
The estimate is based on a number of hypotheses
described in detail in \S\ref{vertical} 
and Appendix~\ref{appb}.
We consider them to be reasonable but the fact that the
vertical velocities are not direct measurements 
must be borne in mind.
The inferred velocities are
insufficient for the penumbral 
radiative flux to be transported by convective motions,
which requires values of the order of 1 km~s$^{-1}$.
However, the finite spatial resolution 
leads to underestimating the true
velocities, a bias whose existence is 
indicated by various results. 
In particular, the same estimate of vertical 
velocities applied to non-magnetic regions
also leads to vertical velocities  of a few hundred m~s$^{-1}$,
although we know from numerical simulations and  
observed Doppler shifts that
the intrinsic granular velocities are much larger
\citep[e.g.,][]{stei98,bec81}.

The algorithm used to infer the 
presence and properties 
of the cork filaments depends   on
several free parameters, e.g., the 
size of the LCT window, the cadence,
the time span of the cork movie, and so on.
They were originally set by trial an error. 
In order to study the
sensitivity of our results on them,
the computation was repeated many times
scanning the range
of possible free parameters (Appendix~\ref{robust}).
This test 
shows how 
the presence of converging flows associated
with dark lanes and downflows
is a robust result, which does not depend on subtleties
of the algorithms.  
It seems to depend on the spatial resolution of the
observation, though.

The downward motions that we find may correspond 
to the ubiquitous downflows indirectly inferred
by \citet{san04b}
from the spectral line asymmetries 
observed in penumbrae.
They may also be 
connected with an old observational result by 
\citet{bec69c},
where they find a local correlation between 
brightness and Doppler shift with
the same sign all over the penumbra, and so,
corresponding to vertical velocities
\citep[see also,][]{san93b,joh93,sch00b}.
The correlation is similar to that characterizing
the non-magnetic granulation,
which 
also suggests
the presence of downflows
in penumbrae.
Analyses of spectroscopic and 
spectropolarimetric sunspot data
indicate the presence of downflows in the
external penumbral rim \citep[e.g.,][]{rim95,
sch00,del01,bel03b,tri04}.
The association between downflows and dark features
is particularly clear in the 0\farcs 2
angular resolution SST data studied by 
\citet{lan05}. 
Again, these downflows may be an spectroscopic
counterpart of those that we infer.
However, it should be clear that we also
find downflows in the inner penumbra,
where they do not.
Whether this fact reflects a true
inconsistency or can be 
easily remedied is not yet known.

\acknowledgements

Thanks are due to G. Scharmer and the SST 
team for granting access to the 
data set employed here.
G. Scharmer also made some valuable comments
on the manuscript. 
The SST is operated by the Institute
for Solar Physics, Stockholm, at the Observatorio
del Roque de los Muchachos of the Instituto  de Astrof\'isica
de Canarias (La Palma, Spain). 
The work has partly been funded by the Spanish Ministry of Science
and Technology, 
project AYA2004-05792,  as well as by
the EC contract HPRN-CT-2002-00313.
%
%
\appendix
\section{Robustness of the results}\label{robust}

The various free parameters involved in the determination
of proper motions
were tunned by trial and error to favor the formation
of narrow  chains of corks in the dark
penumbral lanes.
However, the formation of such chains, their association
with dark features, and the coincidence
with locations of negative horizontal divergence
are all robust results.  We repeat the computation
of proper motions and
cork filament formation varying the free parameters.
The results remain. This section gives a brief account
of the study. 

We start from the {\em reference case}, whose
properties have been described in the main text
(see also the first row in Table~\ref{table1}). 
Then each one of the parameters characterizing this reference
case is varied maintaining the rest unchanged.
The variation modifies velocities,
intensities and angles, a change that we
parameterize using the mean and standard
deviation of the corresponding histograms.  
The actual tests are described
below. Each one is associated with
one or  more rows in Table~\ref{table1}. 
The number
in the first column of Table~\ref{table1}
facilitates the identification; it
corresponds to the number in the list below.
Such cross-reference is needed to  
follow some of the arguments.

%
\begin{deluxetable}{clccccccc}
\tablewidth{0pt}
\tablecaption{Summary of uncertainties\tablenotemark{a}}
\tablehead{\colhead{\tablenotemark{b}}&
	\colhead{Case}\hfill&
	\multicolumn{2}{c}{$U_h$ [km~s$^{-1}$]\tablenotemark{c}}&
	\colhead{$\Delta\mu_I/\sigma_I$\tablenotemark{d}}&
	\multicolumn{2}{c}{$\theta~[^\circ]$\tablenotemark{e}}&
	\multicolumn{2}{c}{$U_z$ [km~s$^{-1}$]\tablenotemark{f}}\\
	&\colhead{}&
	\colhead{start}&\colhead{end}&
	\colhead{}&
	\colhead{start}&\colhead{end}&
	\colhead{start}&\colhead{end}}
\tablenotetext{a}{The figures after
	the $\pm$ signs do not represent error bars 
	but standard deviations of the
	histograms.
	Start and end indicate properties at the
	beginning and the end of the cork movie.
	The empty entries correspond to
	values necessarily identical to the
	reference case. 
	}
\tablenotetext{b}{Number corresponding to the
	list in Appendix~\ref{robust}.}
\tablenotetext{c}{As defined in equation~(\ref{myuh}).
	}
\tablenotetext{d}{As defined in equation~(\ref{shift}).}
\tablenotetext{e}{As defined in equation~(\ref{mytheta}).}
\tablenotetext{f}{As defined in equation~(\ref{myuz}).}	
\tablenotetext{g}{LCT 0\farcs 2~FWHM, penumbra within the
	box in Fig.~\ref{cork1}a, cork movie ends at 110~min, 
	running box 1\farcs 7 wide removed, time step 22~s,
	velocities larger than 3.8~km~s$^{-1}$ neglected.}
\tablenotetext{h}{FWHM of the window used by the LCT algorithm.}	
\tablenotetext{i}{Including the portion of penumbra
	outside the box in Fig.~\ref{cork1}a.}
\tablenotetext{j}{Using a time step of 44~s.}
\tablenotetext{k}{Width of the running box average 
	subtracted to the intensity image.}
\tablenotetext{l}{Diameter of an ideal telescope 
	used to degrade the resolution of the
	original data.}
	
\startdata
&reference\tablenotemark{g} &
	0.51$\pm$0.42&0.21$\pm$0.35&
	-0.21&
	91$\pm$40&89$\pm$42&
	0.00$\pm$0.39&-0.20$\pm$0.36	
	\\
\ref{case1}&
LCT 0\farcs 12\tablenotemark{h}&
	0.57$\pm$0.48& 0.23$\pm$0.41&
	 -0.25&
	91$\pm$40&89$\pm$42&
	0.00$\pm$0.60&-0.44$\pm$0.66
	 \\
\ref{case1}&
LCT 0\farcs 45\tablenotemark{h}&
	0.41$\pm$0.31& 0.19$\pm$0.20&
	 -0.18&
	91$\pm$42&91$\pm$45&
	0.01$\pm$0.17&-0.05$\pm$0.08
	 \\
\ref{case2}&
full penumbra\tablenotemark{i}&
	0.59$\pm$0.50& 0.28$\pm$0.38&
	 -0.16&
	90$\pm$41&91$\pm$43&
	0.00$\pm$0.42&-0.16$\pm$0.38	
\\	
\ref{case3}&
end time 37~min&
	& 0.26$\pm$0.30&
	-0.12&
	&93$\pm$45&
	&-0.14$\pm$0.32
\\
\ref{case3}&
end time 220 min\tablenotemark{j}&
	&0.11$\pm$0.23&
	-0.28&
	&83$\pm$40&
	&-0.24$\pm$0.26
\\
\ref{case4}&
sharpened to 1\farcs 3\tablenotemark{k}& 
	&&
	-0.20&
	91$\pm$40&90$\pm$42&
	&
\\
\ref{case4}&
sharpened to 4\farcs 1\tablenotemark{k}&
	&&
	-0.20&
	91$\pm$41&89$\pm$43&
	&	
\\
\ref{case5}&
no velocity threshold&
	0.58$\pm$0.82& 0.22$\pm$0.54&
	-0.19&
	91$\pm$40&89$\pm$42&
	0.00$\pm$0.83&-0.19$\pm$0.71	
\\
\ref{case10}&
$D$ = 97~cm\tablenotemark{l}
& 0.48$\pm$0.38&0.16$\pm$0.23&
	-0.11&
	90$\pm$41&95$\pm$46&
	0.00$\pm$0.32&-0.21$\pm$0.24
\\
\ref{case10}&
$D$ = 80~cm\tablenotemark{l}
& 0.48$\pm$0.37&0.16$\pm$0.23&
	-0.08&
	90$\pm$41&92$\pm$46&
	0.00$\pm$0.31&-0.22$\pm$0.29
\\
\ref{case10}&
$D$ = 50~cm\tablenotemark{l}
	& 0.47$\pm$0.35&0.15$\pm$0.19&
	-0.01&
	90$\pm$41&94$\pm$47&
	0.00$\pm$0.30&-0.24$\pm$0.25
\\
\ref{case_coor}&
Solar coordinates
	& 0.52$\pm$0.49&0.26$\pm$0.48&
	&
	&&
	0.01$\pm$0.40&-0.20$\pm$0.36
\\
\ref{case_dest}&
No destretching
	& 0.55$\pm$0.43&0.20$\pm$0.32&
	-0.28&
	0.91$\pm$39&0.91$\pm$44&
	0.00$\pm$0.41&-0.28$\pm$0.44
\enddata
\label{table1}
\end{deluxetable}

\begin{enumerate}
\item We change the size of the LCT window to find out that
	the smaller
	the window the clearer the association of
	cork filaments with dark features and 
	downflows. However, such association remains 
	even for windows as large as 0\farcs 45~FWHM.
	\label{case1}

\item The reference case analyzes the portion of
	penumbra within the box in Fig.~\ref{cork1}a.
	When the whole FOV  is used, then the cork
	filaments are not so dark and the mean downflows
	no so intense. However, all trends
	remain.
	\label{case2}
\item We choose the cork filaments
	as they appear after 110~min of evolution of the
	cork movie. This final time is not critical.
	The qualitative properties of the reference
	case are present right from the beginning of the movie,
	and they  are enhanced as time increases.
	Table~\ref{table1} includes means and standard
	deviations at two times, 37~min and 220~min. The latter
	has been computed assuming a  time step of
	44~s to integrate the cork movie.
	 This fact 
	also allows us to discard any significant effect
	of the time step on the results.
	\label{case3}	

\item Changing the size of the box used to remove the
	local mean intensities changes the
	intensities used in our argumentation.
	However, the dependence is very moderate
	as attested by the figures in Table~\ref{table1}.
	\label{case4}
\item
The LCT algorithm gives  a few
velocities larger  than 2 pixels per time step 
(equivalent to 3.8~km~s$^{-1}$), which is 
the largest displacement used to 
compute the cross-correlation
function (\S~\ref{observations}). 
These large velocities 
result from a malfunction
of the algorithm used to determine the
optimum displacement. For this reason
the reference case does not include
velocities larger than  3.8~km~s$^{-1}$.
We checked that this threshold
does not modify our results in  significant way.
We repeated the histograms for $U_h$, $\theta$, $I$
and $U_z$ 
including velocities larger
than 3.8~km~s$^{-1}$.
The means  of the
histograms
do not vary by more than 20~\% (row number \ref{case5} in
Table~\ref{table1}).
\label{case5}
\item 	We investigate the role of the
angular resolution of the images used
to compute the proper motions and 
intensities. The 0\farcs 12 angular resolution
of the original time series was degraded as if
it had been observed with an ideal telescope 
of diameter $D$. (Each snapshot of the
series was convolved 
with the corresponding Airy disk.)
The association between cork filaments and
dark lanes shows up in the
histograms only when the resolution of the degraded
images is not far from  the original one. 
It is almost gone when $D=50~$cm 
($\Delta\mu_I/\sigma_I\sim -0.007$). 
Table~\ref{table1} also includes the cases 
$D=80~$cm and $D=97~$cm. 
(Although the SST
has $D=97~$cm, the original images were restored
so that an ideal $D=97~$cm telescope  does
indeed reduce the contrast of the 
small structures 
used by the LCT algorithm to track.)
It seems that the association between convergent
proper motions and dark lanes depend critically
on the angular resolution. This may explain why
it has remained unnoticed  in previous
studies based on $D=50~$cm class telescopes.
\label{case10}

\item
	We assume the horizontal and vertical
	velocities to be horizontal and vertical
	in a solar coordinate system. However
	the sunspot is slightly out of the disk
	center so that this assumption is
	only approximate. In order to evaluate
	the influence of this approximation, 
	the three components of the 
	velocity field were transformed to
	the true solar horizontal and vertical
	directions. (Two rotations  are needed.) 
	The transformation does not modify the
	histograms of velocities
	in a significant way; see Table~\ref{table1}. 
	\label{case_coor}

\item
	The full analysis was repeated using the
	time series before destretching. The
	different snapshots were co-aligned
	assuming a rigid shift plus the rotation
	to be expected from the alt-azimuthal
	mounting of the SST. This new analysis does
	not differ in a significant way from 
	reference case, so that the histograms
	seems to be independent of the application
	of a destretching algorithm to the data set.
	\label{case_dest}
\end{enumerate}

\section{Derivation of equation~(7)}\label{appb}
This appendix provides a full derivation of equation~(\ref{myuz})
in the vein of the original derivation by \citet{nov89}. 
We contribute by considering the finite spatial resolution
of the observations, and by discussing the validity of the
approximation in the context of the magnetized 
penumbral plasma. 
Such a derivation was found
to be necessary by various readers of the original 
manuscript, including the referees.
The hypotheses are summarized in the final paragraph.

The  continuity equation for a magnetized
plasma with stationary flows is,
\begin{equation}
\nabla (\rho{\bf U})=0,
\label{appb1}
\end{equation}
with the symbols $\rho$ and {\bf U} standing
for the density and
the velocity vector, respectively. 
The observations
provide a kind of volume average of the velocities and
densities, therefore, in order to 
apply equation~(\ref{appb1}) to observables,
the variables $\rho$ and ${\bf U}$
have to be replaced with volume
averages.
We model the volume averages as the convolution with
a kernel describing the 3D region contributing to the 
observed signals.
Using the property
that partial derivatives and convolutions commute
\citep[e.g.][ \S~2]{san96}, 
\begin{equation}
\langle\nabla (\rho{\bf U})\rangle=\nabla 
\langle\rho{\bf U}\rangle=0,
\label{appb3}
\end{equation}
where the angle brackets denote volume
average.
In general,
\begin{equation}
\langle\rho {\bf U}\rangle=
\langle\rho\rangle
\langle{\bf U}\rangle +
\langle
(\rho -\langle\rho\rangle)
({\bf U} -\langle{\bf U}\rangle)
\rangle .
\end{equation}
We will assume that the cross-correlation 
between the local fluctuations of density
in the resolution element 
$\rho -\langle\rho\rangle$,
and the fluctuations of velocity
${\bf U} -\langle{\bf U}\rangle$, 
are smaller than the product of the
mean values, i.e.,
\begin{equation}
|\langle
(\rho -\langle\rho\rangle)
({\bf U} -\langle{\bf U}\rangle)
\rangle|
\ll
|\langle\rho\rangle
\langle{\bf U}\rangle|.
\label{appb2}
\end{equation}
Then,
\begin{equation}
\langle\rho {\bf U}\rangle\simeq
\langle\rho\rangle
\langle{\bf U}\rangle,
\end{equation}
and using equation (\ref{appb3}),
\begin{equation}
\nabla (\langle\rho\rangle\langle{\bf U}\rangle)\simeq 0.
\label{appb4}
\end{equation}
The inequality~(\ref{appb2}) holds if the
fluctuations of density are negligible, if the
fluctuations of velocity are negligible, or
if the fluctuations of density and
velocity are uncorrelated. It is unclear whether 
any of these hypotheses alone justify the
application of equation~(\ref{appb2}) to 
penumbrae. However, the three of them  
can cooperate to make the cross correlation
negligible, so it is not unreasonable to
use equation~(\ref{appb2}) as a working
hypothesis, even for penumbrae.

The  continuity equation for the actual 
variables (equation~[\ref{appb1}]) and the mean 
variables  (equation~[\ref{appb4}])
are formally identical.  We will use the former for the
sake of clarity, dropping the angle brackets
from the expressions
(${\bf U}\equiv \langle{\bf U}\rangle$,
and $\rho\equiv \langle\rho\rangle$). However, it must 
be clear that the densities and velocities 
appearing in all forthcoming equations
correspond to the mean velocities and mean 
densities when the actual velocities and
densities are averaged over a volume equivalent to
the spatial resolution of the observations.

The density of the penumbral plasma 
presents two distinct types of variation. First,
it varies depending on the magnetic field
and temperature, with 
overdensities and underdensities that tend to 
cancel when computing the mean volume averaged
density.
This kind of variation is the only one existing
in horizontal planes. 
Consequently, the horizontal variations of the mean
density are much smaller than the true horizontal
variations of density.
Second, there is a 
systematic drop of density 
with the height in the atmosphere, as required
for a plasma trying to be in  
equilibrium in a strong gravitational field.
It is a systematic effect affecting all magnetic
fields and temperatures and, therefore, it does 
not cancel in the mean density, which is expected
to have a significant vertical variation.
In mathematical parlance, the mean density 
at the spatial coordinates $x$, $y$ and $z$ 
can be
split into two components, 
\begin{equation}
\rho(x,y,z)=\rho_0(z)+\rho_1(x,y,z),
\end{equation}
with the one describing the systematic vertical variation
$\rho_0$ larger than the other accounting for the
horizontal variations $\rho_1$,
 \begin{equation}
\rho_1\ll\rho_0.
\end{equation}
Keeping in mind this condition,
the  continuity equation can be approximated as,
\begin{equation}
\nabla \big[(\rho_0+\rho_1){\bf U}\big]\simeq 
\nabla (\rho_0{\bf U})=
(\nabla\rho_0)\cdot{\bf U}+\rho_0\nabla{\bf U}\simeq 0.
\label{appb5}
\end{equation}
Since $\rho_0$ only varies along the vertical
direction then,
\begin{equation}
(\nabla\rho_0)\cdot{\bf U}={{d\rho_0}\over{dz}}\,
U_z,
\end{equation}
and equation~(\ref{appb5}) can be rewritten as,
\begin{equation}
{{\partial U_z}\over{\partial z}}+
{{d\ln\rho_0}\over{dz}}\,U_z+
{{\partial U_x}\over{\partial x}}+
{{\partial U_y}\over{\partial y}}
\simeq 0,
\label{difeq}
\end{equation}
where the symbols $U_x$, $U_y$ and $U_z$ 
represent the three Cartesian components of the mean
velocity field.
With the density scale height $h_\rho$ defined
as,
\begin{equation}
h_\rho^{-1}=-{{d\ln\rho_0}\over{dz}},
\label{mydef2}
\end{equation}
and the vertical velocity scale height
given by, 
\begin{equation}
h_u^{-1}=-{{\partial\ln|U_z|}\over{\partial z}},
\label{mydef}
\end{equation}
equation~(\ref{difeq}) can be expressed as,
\begin{equation} 
U_z\simeq h_z\,\Big[{{\partial U_x}\over{\partial x}}+
{{\partial U_y}\over{\partial y}}\Big],
\label{myderiv}
\end{equation}
where the symbol $h_z$ represents the scale height of 
the vertical flux of mass,
\begin{equation}
h_z^{-1}=-{{\partial\ln|\rho_0 U_z|}\over{\partial z}}=
h_\rho^{-1}+h_u^{-1}.
\label{tuderiv}
\end{equation} 
The expression~(\ref{myderiv}) corresponds to
equation~(\ref{myuz}), which completes its derivation.
Note that neither $h_\rho$ nor $h_u$ have been assumed 
to be constant, therefore, the densities and velocities 
do not necessarily have to vary exponentially.

An expression formally identical to 
equation~(\ref{myderiv}) is readily 
derived from equation~(\ref{appb1}) under the
unrealistic assumption of constant density. 
Then $\nabla{\bf U}=0$, which leads to  
equation~(\ref{myderiv}) with $h_z=h_u$.
The inference of downflows associated with 
converging flows is based on the assumption 
$h_z > 0$ (see \S~\ref{vertical}).  
In the unrealistic case of constant density,
$h_z > 0\Leftrightarrow h_u > 0$, casting 
serious doubts on the inferred downflows 
since they follow fron the 
sign of $h_u$, which is unknown. 
However, the density is not constant
and, consequently, one can have $h_z > 0$ with  
vertical velocities whose magnitude 
$|U_z|$ decreases with height 
($h_u > 0$), increases with height ($h_u < 0$),
or is constant ($|h_u|\rightarrow\infty$).
Although $h_z > 0$ is an hypothesis,
it seems to be more reasonable than the
alternative $h_z < 0$. The density is expected 
to decrease with  height ($h_\rho > 0$) and  
such a drop  is inherited by the flux of mass,
favoring $h_z > 0$. In other words, for
the assumption $h_z < 0$ to hold one needs flows
of unrealistic amplitude.  Consider the
following example. According to the
arguments in the main text,
$h_\rho\simeq 150$~km. 
Assume  $h_z=-100$~km, so that the upflows and downflows 
are  reversed with respect to those
in Figure~\ref{usubz}. Then equation~(\ref{tuderiv})
renders $h_u=-60$~km, and so, 
\begin{equation}
U_z(z)\simeq U_z(0)\,\exp(z/60\,{\rm km}).
\end{equation}	
This law predicts large unobserved  flows as soon as one  moves up 
in the atmosphere. In the mid photosphere 
with $z=$150\,km,  an upflow  of 0.5 km~s$^{-1}$ at $z=0$~km is
amplified to $U_z$(150 km)=6~km~s$^{-1}$.
Such extreme vertical velocities have not been observed
questioning the underlying assumption
$h_z < 0$.  Conversely, the positive $h_z$ used in the paper 
does not  cause such problems; $h_z$=100\,km 
with $h_\rho\simeq 150$~km lead to $h_u= 300$\,km 
and  $U_z$(150 km)= 0.3 km~s$^{-1}$.

To sum up,
the hypotheses leading to equation~(\ref{myuz}) are:
(a)~the mean flows are stationary, (b)~the
cross-correlation 
between the fluctuations of density and velocity
in the resolution element 
are smaller than the product of the
mean values, and (c)~the mean density varies mostly
in the vertical direction.


\end{document}